\documentclass[5p,twocolumn]{elsarticle}
\usepackage{graphicx,latexsym}
\usepackage{dcolumn}
\usepackage{amssymb,amsmath,bm}
\usepackage{subfigure}
\usepackage{braket}
\usepackage{siunitx}
\usepackage{array}
\usepackage{physics}
\usepackage{float}
\usepackage[nopar]{lipsum}
\usepackage{caption}
\usepackage{multirow}
\usepackage{bigstrut}

\biboptions{sort&compress}

\usepackage[T1]{fontenc}

\usepackage[super]{nth}
\newcommand{\brames}[1]{( #1|}
\newcommand{\ketmes}[1]{|#1)}
\newcommand{\angstrom}{\text{\normalfont\AA}}
\usepackage{hyperref}
\hypersetup{
	pdfnewwindow=true,       
	colorlinks=true,         
	linkcolor=blue,          
	citecolor=blue,          
	filecolor=magenta,       
	urlcolor=black           
}

\usepackage[normalem]{ulem}
\def\b#1{\mathbf{#1}}
\def\mb#1{\mbox{\boldmath$#1$}}
\def\nn{\nonumber \\}
\def\sec#1{Sec.\ \ref{#1}}
\def\eq#1{Eq.\ (\ref{#1})}
\def\fig#1{Fig.\ \ref{#1}}
\def\tab#1{Tab.\ \ref{#1}}

\journal{}

\begin{document}
	
	\begin{frontmatter}

		
		
		\title{Temperature dependence of the single photon source efficiency based on QD-cQED}
		
		\author[a1]{Sarbast W. Abdulqadir}
		\address[a1]{Division of Computational Nanoscience, Physics Department, College of Science, 
			University of Sulaimani, Sulaimani 46001, Iraq}
		
		\author[a1]{Hawri O. Majeed}
		
		\author[a1,a2,a3]{Nzar Rauf Abdullah}
		\ead{nzar.r.abdullah@gmail.com}
		\cortext[correspondingauthor]{Corresponding author: Nzar Rauf Abdullah}
		\address[a2]{Computer Engineering Department, College of Engineering, Komar University of Science and Technology, Sulaimani, Iraq}
		\address[a3]{Science Institute, University of Iceland, Dunhaga 3, IS-107 Reykjavik, Iceland}

		
		\begin{abstract}
			
			We study a photonic circuit consisting of a quantum dot, QD, coupled to a photon cavity over a wide range of temperature up to room temperature. A key component of such a system is presented here in the form of a Purcell-enhanced single-photon source based on Cavity Quantum Electrodynamics, cQED. We use a real set of pure dephasing data extracted from experimental measurements of InGaAs QD to calculate the effective QD-cavity coupling strength, the Purcell factor, and the single photon efficiency emerged from the QD-cavity system in the cases without and with detuning. In the non-detuned system, the effective coupling strength between the QD and the resonator decreases with increasing temperature, results in a decrease in efficiency. However, when the temperature of the QD-cavity system increases under Purcell effect conditions, the detuned QD-cavity system induces spontaneous emission rate enhancement. As a result, we found that the increase in efficiency can be obtained under a certain condition, when the maximum effective coupling strength and the Purcell factor are related to the spontaneous emission and the pure dephasing rates. Additionally, the influences of the pumping mechanism on the efficiency of the QD-system were examined and showed that the pumping process can be used to further increase in efficiency. Our results can be advantageous for advanced quantum optics applications once temperature is taken into account. 
		\end{abstract}


	\end{frontmatter}

	\section{Introduction} 
	
	The emerging field of quantum key distribution, quantum repeaters and quantum information science requires the development of a new type of light source, in which a high-quality single-photon source and photon number could be carefully controlled \cite{Reimer2019, 10.1117/12.912969, Senellart2017}.
	In classical picture of light sources, photons consist of a macroscopic number of emitters 
	in which the Poisson or super-Poisson statistics is used to define their distribution \cite{walls2007quantum}, while the fields of quantum theory of light have attempted to produce 
	a single quantum emitter with a photon stream containing one and only one photon in a given time interval \cite{Paesani2020, Tomm2021}. Such as “antibunched” source
	will be greatly important in the new field of quantum cryptography, where security from eavesdropping depends on the ability to generate no more than one photon at a time \cite{PhysRevLett.89.187901, Heindel_2012}.
	
	Among the different sources available to generate a single photon source, also quantum dot coupled to a cavity is one of the attempts \cite{michler2000quantum, doi:10.1063/1.1415346, PhysRevB.71.241304, Heinze2015}. The QD-cavity photon source can allow emission in a well-defined direction and increased emission efficiency of the devices \cite{10.1117/12.543168, PhysRevB.90.155303, Wang2019}. 
	The efficiency of a single photon sources based on QD-cavity devices has been studied by 
	Cui and {\textit{et al.}}, where they demonstrated that the efficient single photon source are obtained if the coupling of the emitter to the single cavity mode is far strong with almost no dephasing of the emitter during the emission process \cite{Cui_05}. It thus keep the single photon emission process to be deterministic, consecutive, and indistinguishabl.
	
	The indistinguishability of photons in QD-microcavity system through a Hong–Ou–Mandel-type two-photon interference experiment has been studied, and a largely indistinguishable of 
	consecutive photons are found \cite{Santori2002}. The degree of indistinguishability of 
	incoherently excited quantum dots in the context of cavity quantum electrodynamics, cQED, is further improved \cite{PhysRevB.100.155420}. Furthermore, an integrated ring resonators coupled to a QD could 
	also be used to enhance the photon indistinguishable degree \cite{Dusanowski_2020}.

	A two level QD-cavity could be integrated more easily and it has been considered to be the most potential single photon source \cite{Dory2016}, while the QD in bulk materials
	are very inefficient at producing single photon \cite{Santori2002}. It has also been shown that the cQED can increase the efficiency of two-level QD single photon source. The QD-cQED can increase direct emission of photons spontaneously in a direction of interest, and
	it improves the dissipation mechanism which enables pure quantum state emission to occur \cite{Alex_2010}.
	In addition, two more physical parameters are important in determining the efficiency of a single photon source which are dephasing process and detuning of QD-cQED. 
	It is demonstrated that the relation between efficiency of single photon source and pure dephasing are inversely proportional to each other in a perfect resonance of QD-cQED (without detuning), but the efficiency is improved by pure dephasing in the detuning QD-cQED \cite{PhysRevA.79.053838}.
	
	There are several factors that can control pure dephasing of two level QD-cQED which are 
	an applied electric field near the two-level QD, the temperature, $T$, and the pump rate of the cQED. These parameters are experimentally controllable \cite{PhysRevB.75.073308}. 
	Based on the aforementioned results, we try to present the temperature dependent of the single photon source efficiency based on QD-cQED, the effective QD-cavity coupling strength, and the Purcell factor. We also consider the effects of pumping on efficiency of the single photon sources. In addition, the influence of internal and external losses on these parameters are shown.
	
	We arranged the current work as follows: in \sec{section_model} the Hamiltonian of the QD-cavity system, and a master equation formalism are demonstrated. In \sec{section_results} the main obtained results are presented. In \sec{section_conclusion}, the  conclusion is shown.

	\section{Theory and Formalism}\label{section_model}
	
\begin{figure}[htb]
		\centering
		\includegraphics[width=0.4\textwidth]{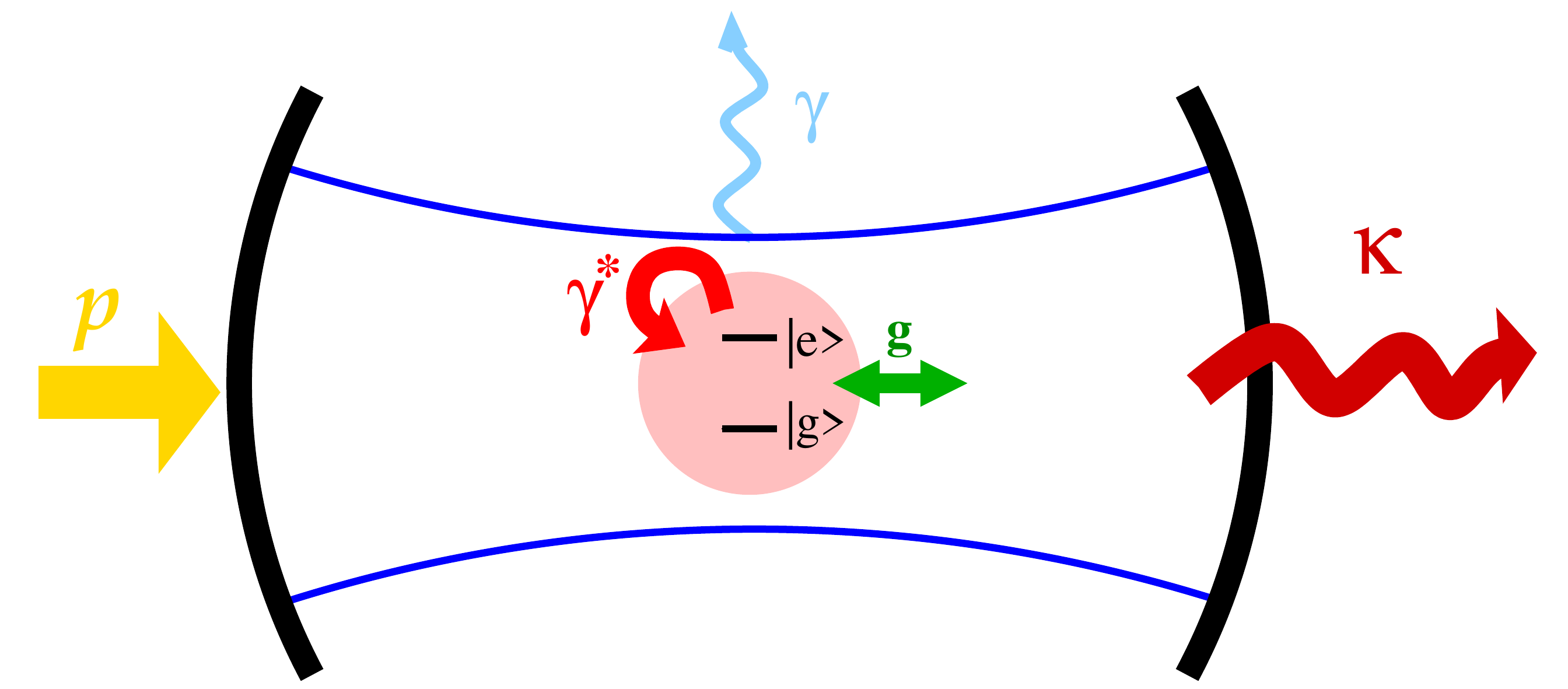}
		\caption{Schematic diagram of a two level QD coupled to a single mode cavity.}
		\label{fig01}
	\end{figure}

	The general method representing a cQED model consists of a two level QD coupled to a cavity with a single mode shown in \fig{fig01}. The total Hamiltonian of the system, the QD and the cavity, is given by \cite{RevModPhys.87.1379, PhysRevB.81.245419},
	
	\begin{equation}
		\hat{H} = \omega_{\rm QD} \, \hat{\sigma}_{+} \hat{\sigma}_{-} 
		+\omega_{\rm c} \, \hat{a}^{\dagger} \hat{a} 
		+ i g \, (\hat{a}^{\dagger} \hat{\sigma}_{-} - \hat{\sigma}_{+} \hat{a}).
	\end{equation}
	
	Where $\omega_{\rm QD}$ is the frequency of two level QD, $\hat{\sigma}_{-}$ and $\hat{\sigma}_{+}$ are the electron lowering and rising operators of the QD, respectively, $\omega_{\rm c}$ displays the photon frequency of the cavity, $\hat{a}^{\dagger}$ and $\hat{a}$ are the photon annihilation and creation operator in the single  mode of the cavity, respectively, and $g$ is the QD-cavity coupling strength. We assume that Planck's constant is one in the calculations, $\hbar=1$.
	The QD and cavity frequencies including the detuning, $\delta$, is defined as $\delta = \omega_{\rm QD} \text{-}\omega_{\rm c}$.
	It is considered that the QD is initially pumped in its excited state. There are also several loss factors in the QD-cavity which are the losses of isolated QD denoted by $\gamma$, cavity decay rate, $k$, the pure dephasing, $\gamma^*$, and the pumping rate to the cavity with strength $\mathcal{P}$ (see \fig{fig01}).
	
	The time evolution of the open QD-cavity can be calculated via 
	the standard approach of density operator, $\hat{\rho}$, in which the Lindblad master equation is used in the below form \cite{RevModPhys.87.1379, ABDULLAH2020114221}
	
	\begin{equation}
		\frac{d\hat{\rho}}{dt} = \frac{1}{i} [\hat{H}, \hat{\rho}] + \mathcal{L}_{\rm diss},
		\label{eq_masterequation}
	\end{equation}
	
	where $\mathcal{L}_{\rm diss}$ is the dissipation terms describing the damping parts due to the QD, $\mathcal{L}_{\rm QD}$, the cavity $\mathcal{L}_{\rm c}$, the dephasing, $\mathcal{L}_{\rm deph}$, and the pumping, $\mathcal{L}_{\rm pump}$, in the Lindblad form which is given by
	
	\begin{equation}
		\mathcal{L}_{\rm diss} = \mathcal{L}_{\rm QD} + \mathcal{L}_{\rm c} + \mathcal{L}_{\rm deph} + \mathcal{L}_{\rm pump}, 
	\end{equation}
	
	herein 
	
	\begin{equation*}
		\mathcal{L}_{\rm QD} = (\gamma/2) \big[ 2 \hat{\sigma}_{-} \hat{\rho} \hat{\sigma}_{+} 
		- \hat{\sigma}_{+} \hat{\sigma}_{-} \hat{\rho}
		- \hat{\rho} \hat{\sigma}_{+} \hat{\sigma}_{-}  \big],
	\end{equation*}
	
	\begin{equation*}
		\mathcal{L}_{\rm c} = (\kappa/2) \big[ 2 \hat{a} \hat{\rho} \hat{a}^{\dagger}
		- \hat{a}^{\dagger} \hat{a} \hat{\rho}
		- \hat{\rho} \hat{a}^{\dagger} \hat{a}  \big],
	\end{equation*}
	
	\begin{equation*}
		\mathcal{L}_{\rm deph} = (\gamma^*/4) \big[ \hat{\sigma}_z \hat{\rho} \hat{\sigma}_z -\hat{\rho} \big],
	\end{equation*}
	
	with $\hat{\sigma}_z = \hat{\sigma}_{+} \hat{\sigma}_{-} - \hat{\sigma}_{-} \hat{\sigma}_{+}$, and 
	
	\begin{equation*}
		\mathcal{L}_{\rm pump} = (\mathcal{P}/2) \big[ 2 \hat{\sigma}_{+} \hat{\rho} \hat{\sigma}_{-} 
		- \hat{\sigma}_{-} \hat{\sigma}_{+} \hat{\rho}
		- \hat{\rho} \hat{\sigma}_{-} \hat{\sigma}_{+}  \big].
		\label{L_pumping}
	\end{equation*}

	In order to obtain the density matrix, $\hat{\rho}$, the master equation, \eq{eq_masterequation}, is solved numerically using Qutip software \cite{JOHANSSON20121760, JOHANSSON20131234}. However, we have used numerical method to solve it as we have done it for quantum transport through quantum dot or wire coupled to a photon cavity  \cite{PhysRevB.82.195325, doi:10.1021/acsphotonics.5b00115, Vidar:ANDP201500298}.
	
	The loss rate of the cavity consists of two parts: The internal loss rate, $\kappa_{\rm in}$, and the external loss rate, $\kappa_{\rm out}$, which give $\kappa = \kappa_{\rm in} + \kappa_{\rm out}$. Both types of losses of the cavity can be experimentally observed \cite{doi:10.1063/1.3527930}. Taking into account the geometrical information of the cavity, the total loss rate of the cavity is defined as 
	$\kappa = \big( \pi \, c \, d^2/8V \big) \times \big[ (1 - \sqrt{R_l R_r})/(R_l R_r)^{1/2} \big]$, and the internal loss rate is $\kappa_{in} = (\pi \, c \, d^2/4 V) \alpha$, where $c$ is speed of light, $d$($V$) displays the diameter(volume) of the cavity, $R_l(R_r)$ is the reflectivity of left(right) mirror of the cavity, $\alpha$ presents one round-trip of the cavity internal loss. In addition, the QD-cavity coupling constant is introduced as $g = \big( M^2 \omega_{\rm QD} / 2\varepsilon_0 \hbar V \big)^{1/2} = \big( M^2 (\delta - \omega_{\rm c}) / 2\varepsilon_0 \hbar V \big)^{1/2}$, where $M$ denotes dipole moment of the two level QD transition, and $\varepsilon_0$ is the permittivity of vacuum \cite{RevModPhys.87.1379}.
	
	The evolution of the electron and photon numbers in the QD-cavity have been well described in \cite{englund2009coherent, PhysRevA.79.053838}. In the case of detuning in the QD-cavity, 
	one can write the evolution of the electron and photon numbers in terms of the effective coupling
	rate $R$ as follows \cite{https://doi.org/10.1002/lpor.200810081, JONSSON201781}: 
	
	\begin{equation}
		\frac{d}{dt} \expval{\hat{\sigma}_{+}\hat{\sigma}_{-}} = 
		R \expval{ \hat{a}^{\dagger} \hat{a}} 
		- (\gamma + R) \expval{\hat{\sigma}_{+}\hat{\sigma}_{-} }, 
	\end{equation}
	
	and 
	
	\begin{equation}
		\frac{d}{dt} \expval{\hat{a}^{\dagger}\hat{a}} = R \expval{\hat{\sigma}_{+}\hat{\sigma}_{-}} 
		- (\kappa + R) \expval{ \hat{a}^{\dagger} \hat{a}}. 
	\end{equation}
	
	Herein, the incoherent regime is considered in which the adiabatically elimination can be performed \cite{7798963, Azouit_2017}. The effective coupling rate or effective transfer rate, $R$, in the absence of the pumping to the cavity, $\mathcal{P} = 0.0$,  is given by
	
	\begin{equation}
		R = \frac{4g^2}{\kappa + \gamma + \gamma^*}  \times  \frac{1}{1 + \Big[  \frac{2\delta}{\kappa + \gamma + \gamma^*} \Big]^2}.
	\end{equation}
	
	Taking into account the photoluminescence, the QD-cavity is pumped with a specific value of pumping strength rate,  $\mathcal{P}$. In this case, the effective coupling rate, $R$, is modified to \cite{PhysRevB.81.245419, GUDMUNDSSON20181672}
	
	\begin{equation}
		R = \frac{4g^2}{\Gamma} \times \frac{1}{1 + (2\delta/\Gamma)^2}.
		\label{R_pumping}
	\end{equation}	
	
	With $\Gamma = \mathcal{P} + \gamma + \gamma^{*} + \kappa$. In this case, the \eq{L_pumping} has to be considered in solving Lindblad master equation.
	
	Once the effective coupling rate, $R$, is known, the efficiency of the corresponding single photon source, $\mathcal{E}$, can be found using 
	
	\begin{equation}
		\mathcal{E} = \frac{\kappa_{\rm out}}{\kappa_{\rm in} + \kappa_{\rm out}} \times \frac{R \big( \frac{1}{\kappa} + \frac{1}{\gamma} \big)}{1 + R\big( \frac{1}{\kappa} + \frac{1}{\gamma} \big)}.
		\label{eq:efficiency}
	\end{equation}
	
	There is a regime of interest for the QD-cavity system called bad cavity system in which $R \ll \kappa$. This corresponds to the so-called Purcell regime and the system is characterized by a generalized Purcell factor $F^{*}$ indicating the enhancement of the spontaneous emission
	rate. The Purcell factor is given by
	
	\begin{equation}
		F^{*} = \frac{4g^2}{\gamma(\kappa + \gamma + \gamma^*)} \times 
		\frac{1}{1 + (\frac{2\delta}{\kappa + \gamma + \gamma^*})^2}.
		\label{Purcell_Factor}
	\end{equation}

	\section{Results}\label{section_results}
	
\begin{figure}[htb]
		\centering
		\includegraphics[width=0.45\textwidth]{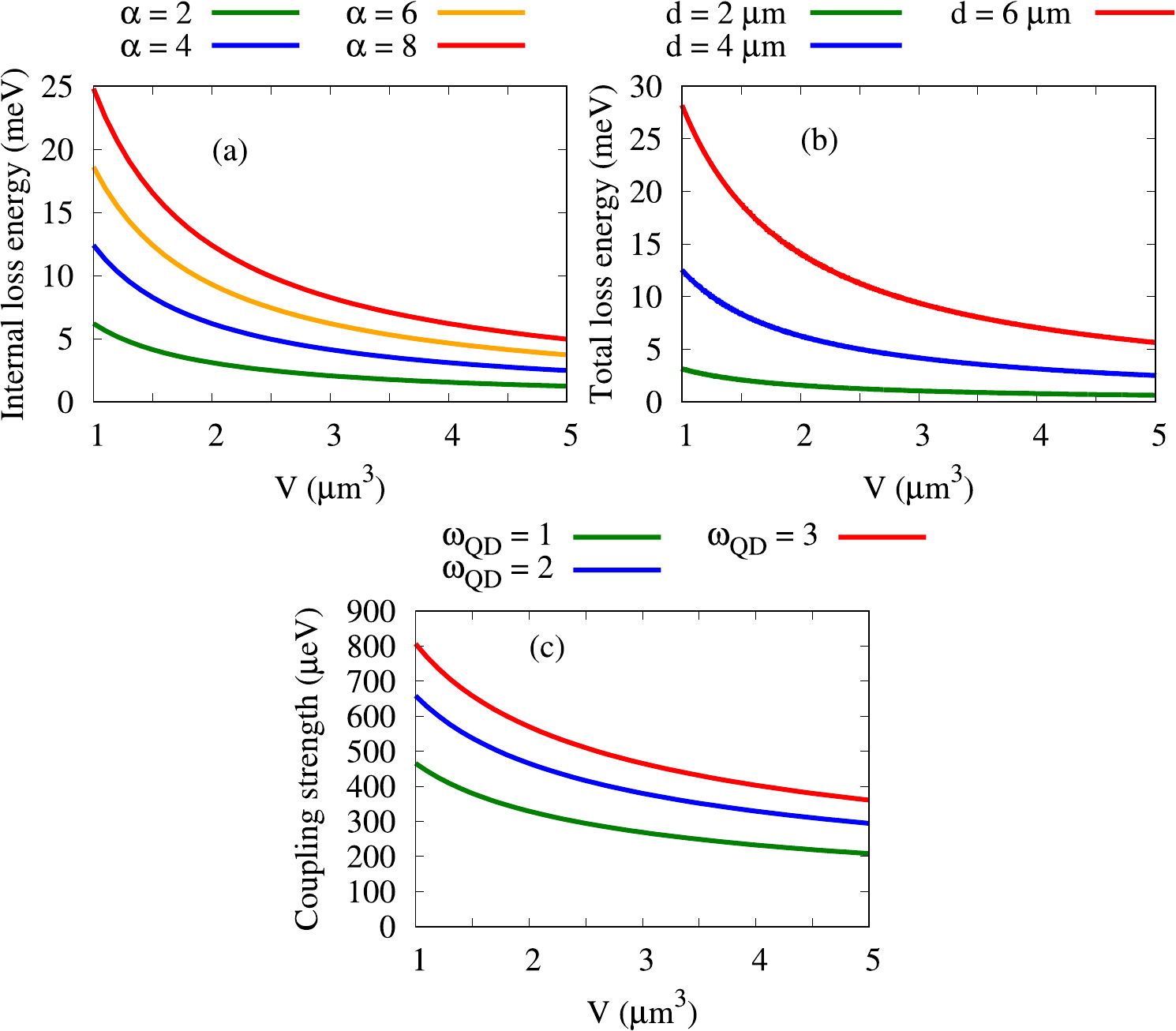}
		\caption{(a) Internal loss energy versus volume for $\alpha = 2$ (green), $4$ (blue), 
			$6$ (orange), and $8$ (red) where $d = 2.0 \, \mu m$.
			(b) Total loss energy as a function of volume for different values of diameter of the cavity, $d = 2$ (green),  $4$ (blue), and $6$~$\mu m$ (red). (c) QD-cavity coupling energy, $\hbar g$, for different values of transition frequency between the two levels of the QD where $\omega_{\rm QD} = 1.0$ (green), $2.0$ (blue), and $3.0$~MHz (red). It is considered that $R_l = R_r = R_0 = 0.99$.}
		\label{fig02}
	\end{figure}

	In this section, we present the main results of the temperature dependent of the effective coupling, the Purcell factor, and efficiency of a single photon source based on InGaAs quantum dot coupled to a cavity.
	We first present geometrical effects of QD-cavity on internal loss energy, $\kappa_{\rm in}$ (a), total loss energy, $\kappa$ (b), and QD-cavity coupling energy, $\hbar g$ (c) in \fig{fig02} in which 
	these physical parameters are plotted versus volume of the cavity, $V$. It is assumed that 
	the reflectivity rate of both sides of the cavity are equal, $R_l = R_r = R_0 = 0.99$.
	Basically, the external loss rate, $\kappa_{\rm out}$, corresponds to the design
	of the cavity input mirror and the internal loss rate, $\kappa_{\rm in}$, could arise from residual
	absorption or diffusion or inelastic scattering through the cavity leaky modes (side leakage) \cite{doi:10.1063/1.3527930}.
	The internal loss and total loss energies are decreased with increasing the cavity volume which 
	indicates a photon in the cavity has a larger facility leading to less interaction with the wall of the cavity. As a result, the residual absorption or diffusion through the cavity leaky modes is decreased, and the loss rates are thus decreased. 
	
	In addition, if $\alpha$ is further increased, the internal total loss rate is enhanced, and the relation between loss and diameter of the cavity, $d$, is proportional as it is expected. It is clear to see that the QD-cavity coupling strength energy is also decreased with increasing the cavity volume for all considered values of $\omega_{\rm QD}$.

\begin{figure}[ht!]
		\centering
		\includegraphics[width=0.35\textwidth]{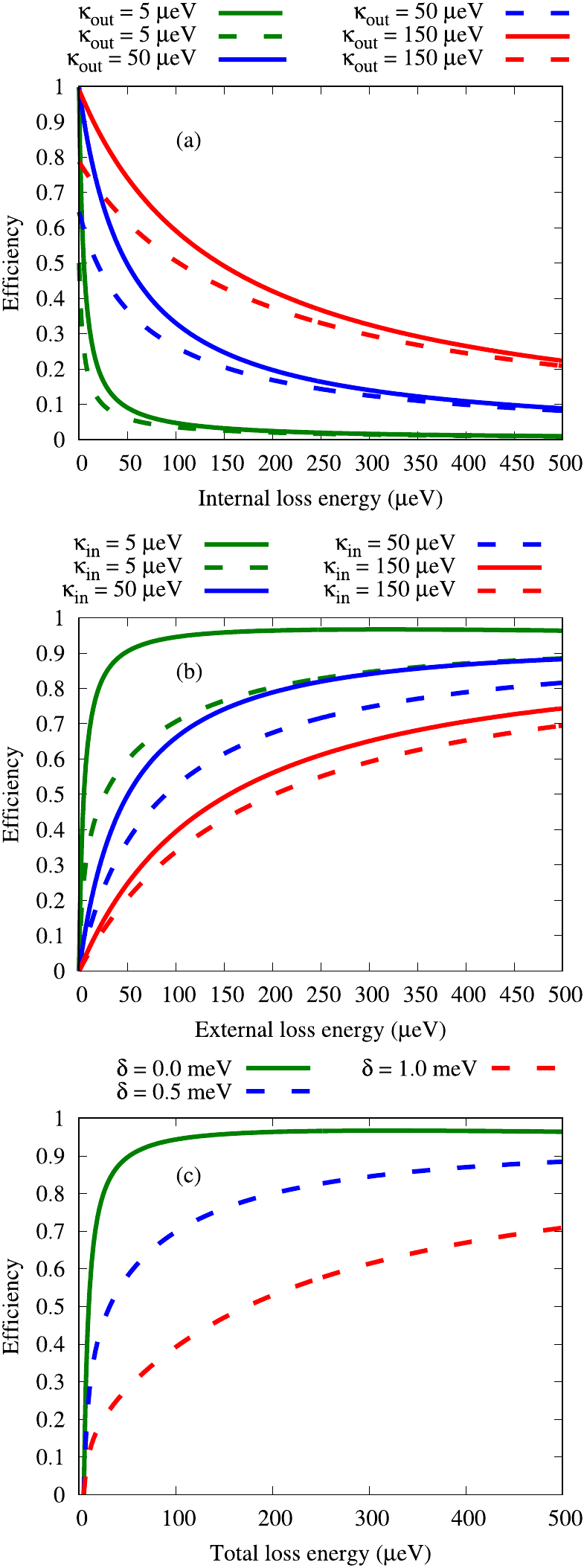}
		\caption{Efficiency as a function of internal loss energy (a), external loss energy (b), and total loss energy (c) for the non-detuning (solid lines) and detuning (dashed line) of the QD-cavity. The QD-cavity coupling energy $g = 50$~$\mu$eV, $\gamma = 0.01 \, g = 0.5$~$\mu$eV, $\gamma^* = 0.04$~meV, and $\delta = 10 \, g$ in both (a, b), while $\delta = 10 \, g = 0.5$~meV (dashed blue), and $\delta = 20 \, g = 1.0$~meV (dashed red) in (c).}
		\label{fig03}
	\end{figure}

	The experimental situations for measuring the effective coupling rate, the Purcell factor, and  the efficiency consider a very small spontaneous emission rate, $\gamma$, comparing to the other typical loss rates such as total loss rate or cavity damping rate, $\kappa$. Based on this fact, 
	we assume that the QD-cavity coupling strength is $g = 50$~$\mu$eV, $\gamma = 0.01 \, g = 0.5$~$\mu$eV, and tune the total loss rate, $\kappa$ in such away that the value of $\gamma$ is much smaller than $\kappa_{\rm in}$, $\kappa_{\rm out}$ and $\kappa$ as they are shown in \fig{fig03}. 
	In the figure, the effects of internal, external, and the total loss rates on the efficiency of the single photon source ($\mathcal{E}$) are also presented.
	
	We have done these calculations at a specific temperature, $T = 50$~K, corresponding to the pure dephasing rate  $\gamma^* = 0.04$~meV for a QD made of InGaAs \cite{PhysRevLett.87.157401}.
	In \fig{fig03}(a), it is shown that the internal loss energy decreases the efficiency of the cavity for the non-detuning (solid lines) and detuning (dashed lines) system. This efficiency reducation refers to the residual absorption or diffusion through the side leakage or micropillar leaky modes.
	With increasing external loss energy representing the energy of the single photon obtained from the QD-cavity system, $\kappa_{\rm out}$, the efficiency is enhanced which can be confirmed via \fig{fig03}(b). We know that the value of the efficiency becomes almost half ($\mathcal{E} = 0.5$) when the internal loss energy is equal to the external loss energy, $\kappa_{\rm in} = \kappa_{\rm out}$.
	
	In \fig{fig03}(c), the efficiency versus the total loss energy, $\kappa$, is plotted for the non-detuning (solid line), and detuning (dashed lines) of the QD-cavity system with strength of $\delta = 10 \, g = 0.5$~meV (dashed blue) and $20 \, g = 1.0$~meV (dashed red), where the internal loss is assumed to be $\kappa_{\rm in} = 5.0 \, \mu$eV. As the detuning parameter is increased, the spontaneous radiation life time increases leading to reducation in the 
	efficiency.

\begin{figure}[htb]
		\centering
		\includegraphics[width=0.35\textwidth]{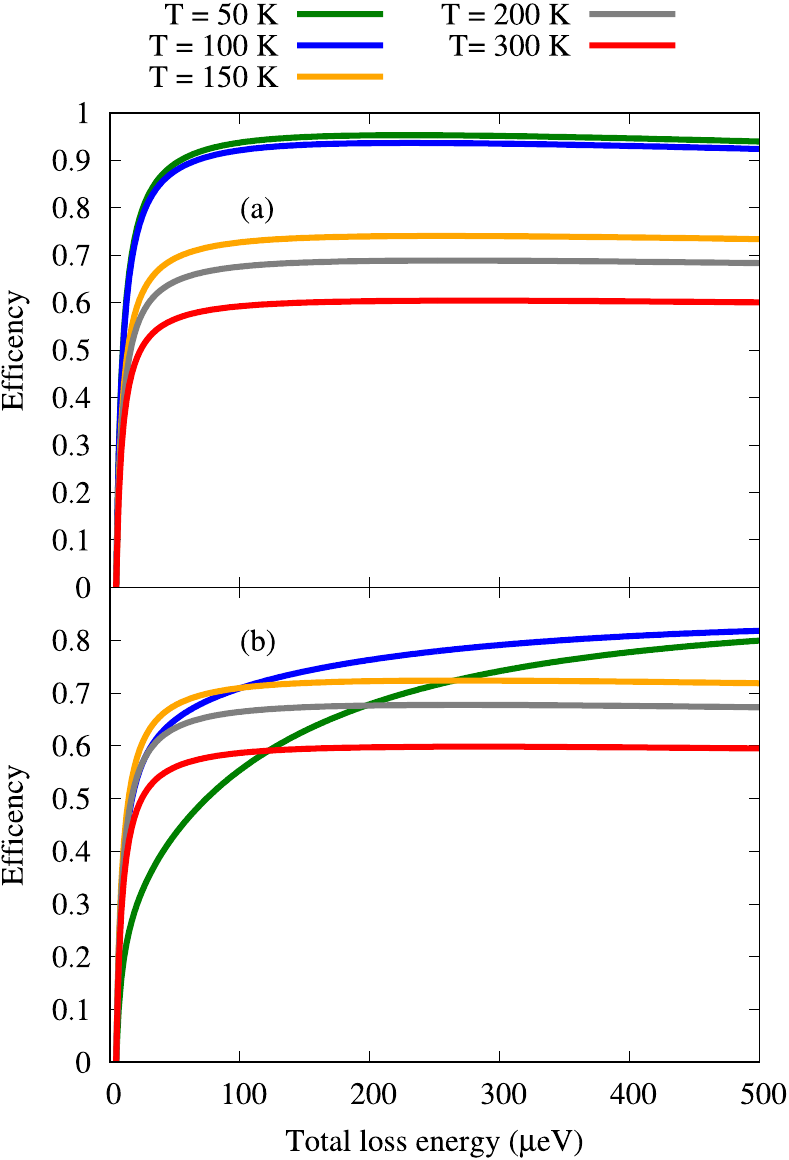}
		\caption{Efficiency as a function of total loss energy for different temperature, $T = 50$ (green), $100$ (blue), $150$ (orange), $200$ (gray), and $300$~K (red). The coupling strength is $g = 50$~$\mu$eV, internal lose rate is $\kappa_{\rm in} = 5$~$\mu$eV, $\gamma = 0.02 \, g$, and detuning parameter is $\delta = 0.0$ (a) and $10 \, g$ (b).}
		\label{fig04}
	\end{figure}

	Several approaches to control the efficiency of a single photon source based on cQED have been used such as electric field \cite{PhysRevLett.103.087405} and temperature \cite{PhysRevB.75.073308, doi:10.1021/acs.nanolett.5b03724}. The temperature influences efficiency of a single-photon source via pure dephasing parameter, $\gamma^*$. 
	In \fig{fig04}, the efficiency as a function of the total lose is plotted for different value of temperature ranging from $50$ to $300$~K for the non-detuning (a) and detuning (b) of the QD-cavity system where the detuning rate is $\delta = 10 \, g = 0.5$~meV.  
	We should mention that the state of the art parameters of QD-cavity are used here in which the coupling strength is $g = 50$~$\mu$eV, internal lose rate is $\kappa_{\rm in} = 5.0$~$\mu$eV, $\gamma = 0.02 \, g = 1.0$~$\mu$eV, and $\delta = 10 \, g$ in \fig{fig04}. In addition, we have selected five different temperature, $50$, $100$, $150$, $200$, and $300$~K corresponding to the pure dephasing parameter $\gamma^* = 0.04$, $0.22$, $3.0$, $4.0$, and $6.0$~meV, respectively, for the InGaAs QD system \cite{PhysRevLett.87.157401}. It is interesting to see that the efficiency is decreased with increasing the temperature in the non-detuning QD-cavity, while an enhancement in efficiency is seen for the detuning system as the temperature is increased up to $100$~K.
	
\begin{figure}[htb]
		\centering
		\includegraphics[width=0.45\textwidth]{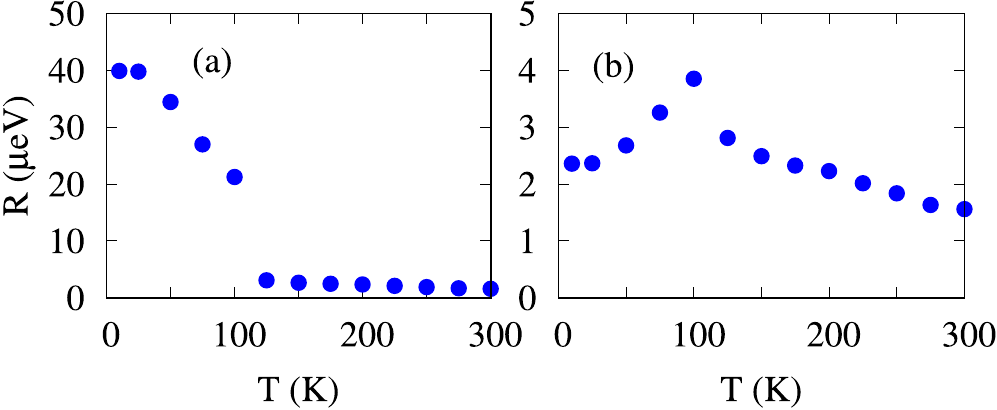}
		\caption{Effective transfer rate, $R$, as a function of temperature for the
			non-detuning (a) and detuning, ($\delta = 10 \, g$) (b), of the QD-cavity system,  where $\gamma = 0.02 \, g$, $g = 50$~$\mu$eV, and $\kappa = 5 \, g$.}
		\label{fig05}
	\end{figure}

	To get insight into understanding the temperature-dependent efficiency, the effective transfer rate, $R$, is plotted as a function of temperature for the non-detuning (a) and detuning (b) of the QD-cavity in \fig{fig05}. We have selected several measured values of pure dephasing (temperature) for the InGaAS QD obtained in \cite{PhysRevLett.87.157401}.
	First, the effective transfer rate for the detuning system is ten times smaller than that of the non-detuning system for the temperature below $100$~K. This leads the efficiency of the detuning system (see \fig{fig04}(b)) to be smaller than that of the non-detuning system for the temperature below $100$~K. Furthermore, the effective transfer rate seems to be similar for the detuning and non-detuning systems when the temperature is above $100$~K which gives rise very similar efficiency for detuning and non-detuning systems above $100$~K. 
	Another difference between detuning and non-detuning systems is that the effective transfer rate is decreased for the detuning and increased for non-detuning system in temperatures below $100$~K. 
	We therefore see maximum efficiency for detuning QD-cavity system at $100$~K as the effective transfer rate is maximum at this temperature.
	
	It has been reported that the maximum enhancement of effective transfer rate, $R$, for the detuning QD-cavity system is obtained if $\kappa + \gamma + \gamma^* \approx \delta$ allowing to reach an optimal value of $R$ which is $R_{\rm max} \approx g^2/\delta$. Using the selected values of $\gamma$, $\gamma^*$, and $\kappa$ that are used in our calculations, we obtain the maximum value of $R$ which is $R_{\rm max} = 4.0$~$\mu$eV at $T = 100$~K corresponding to the pure dephasing, $\gamma^* = 0.22$~meV as it is shown in \fig{fig05}(b).
	
\begin{figure}[htb]
		\centering
		\includegraphics[width=0.45\textwidth]{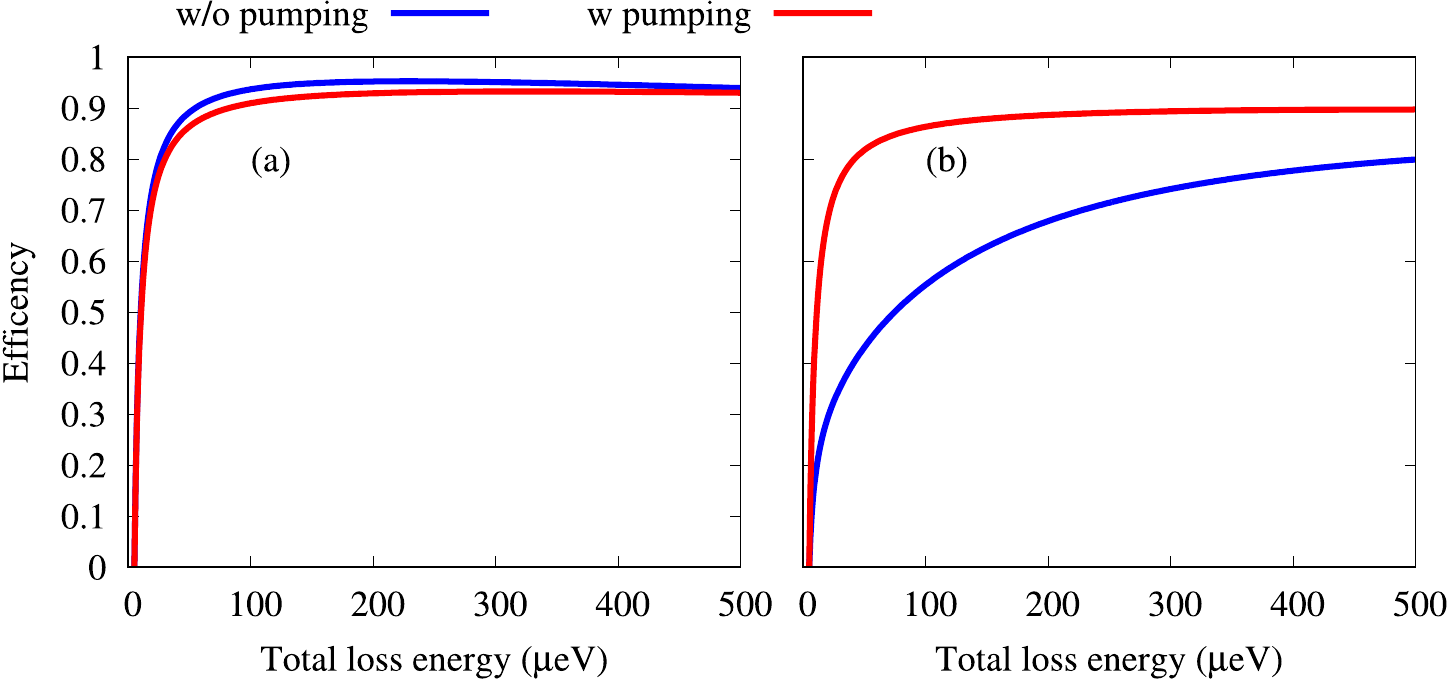}
		\caption{Efficiency as a function of total loss energy for the system without (w/o), $\mathcal{P}=0.0$ , and with (w) pumping, $\mathcal{P}=2 \, g$, in the case of the non-detuning (a) and  detuning, ($\delta = 10 \, g$) (b) of the QD-cavity system, where $T = 50$~K, $\gamma^* = 0.04$~meV, $\gamma = 0.02 \, g$, $g = 50$~$\mu$eV, and $\kappa = 5 \, g$.}
		\label{fig06}
	\end{figure}

	The QD-cavity system can be used to study usual photoluminescence experiment where the QD is typically pumped in a continuous wave \cite{Ishida2013}. During the pumping process, we have to take the dissipation term into account presented in \eq{L_pumping}. 
	The pumping mechanism also influences the efficiency of the single photon via the effective transfer rate, $R$, presented in \eq{R_pumping}. We now consider the QD-cavity system without, w/o, (blue) and with pumping (red) for the non-detuning (a) and detuning (b) QD-cavity system shown in \fig{fig06}, where the pure dephasing is $0.04$~meV corresponding to temperature $T = 50$~K. An intermediate pumping strength is considered in our calculations where the pumping strength is  $\mathcal{P} = 2.0 \, g$ which is approximately equal to $\gamma + \gamma^* + \kappa$.
	The pumping process slightly decreases the efficiency of the single photon source in the non-detuning system while it prominently enhances the efficiency in the detuning system. 
	We can see from \eq{R_pumping} that the effective coupling rate is inversely proportional to pupmping strength, $\mathcal{P}$, in the case of non-detuning system, $\delta = 0.0$. Increasing the $\mathcal{P}$ will decrease the effective coupling rate and thus the efficiency is reduced. 
	In contrast, it is clearly seen that the relation between $\mathcal{P}$ and $R$ is directly proportional at low value of pure dephasing for the detuning system as the second term of $R$ in \eq{R_pumping} is proportional to $\mathcal{P}^2$.  We therefore see an enhancement in the efficiency for the pumped detuning QD-cavity system.

\begin{figure}[htb]
		\centering
		\includegraphics[width=0.4\textwidth]{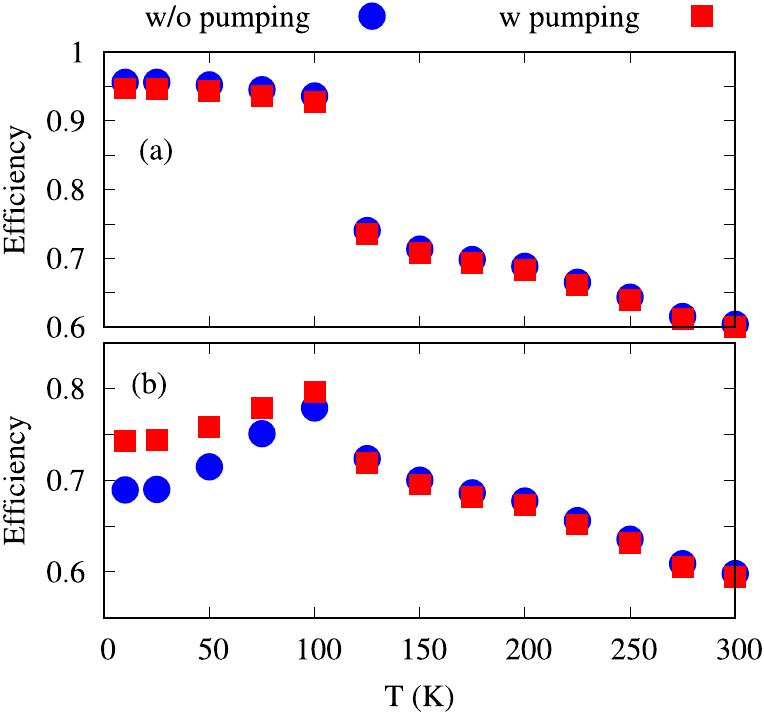}
		\caption{Efficiency as a function of total loss energy for the system without (w/o), $\mathcal{P}=0.0$ , and with (w) pumping, $\mathcal{P}=2 \, g$, in the case of the non-detuning (a) and detuning, ($\delta = 10 \, g$) (b) of the QD-cavity system, where $T = 50$~K, $\gamma^* = 0.04$~meV, $\gamma = 0.02 \, g$, $g = 50$~$\mu$eV, and $\kappa = 5 \, g$.}
		\label{fig07}
	\end{figure}

\begin{figure}[htb]
		\centering
		\includegraphics[width=0.4\textwidth]{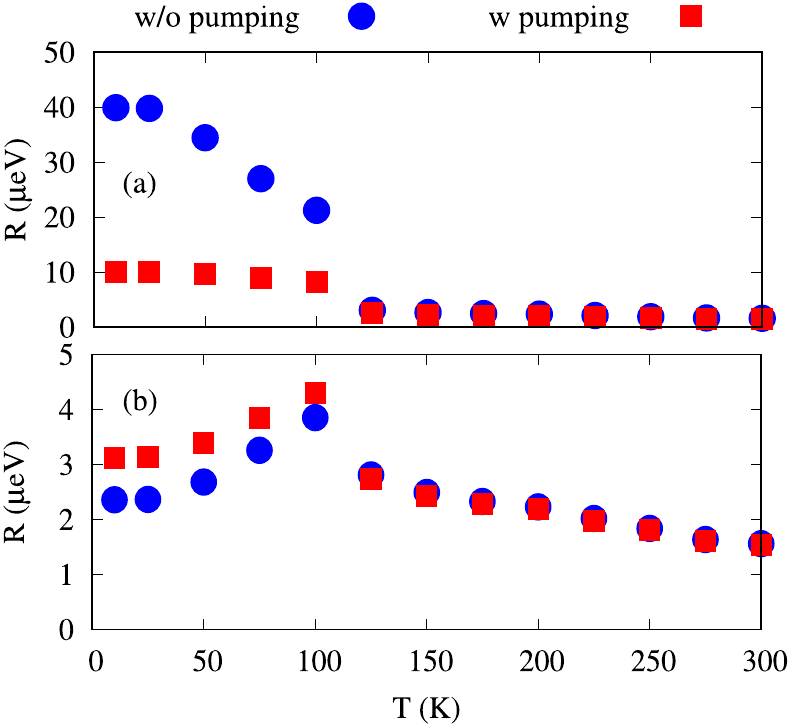}
		\caption{Effective transfer rate, $R$, as a function of temperature for the
			non-detuning (a) and detuning, ($\delta = 10 \, g$) (b) of QD-cavity system in the case of without pumping, $\mathcal{P}=0.0$ (blue), and with pumping, $\mathcal{P}=2 \, g$ (red), where $\gamma = 0.02 \, g$,
			$g = 50$~$\mu$eV, and $\kappa = 5 \, g$.}
		\label{fig08}
	\end{figure}

	Our next goal is to study the QD-cavity system with (w) and without (w/o) pumping in a wider range of pure dephasing proportional to the temperature. 
	To further see the effects of pumping process on efficiency, we demonstrate the efficiency plotted for a wide range of temperature in \fig{fig07} where the temperature is taken from
	$10$ to $300$~K for the non-detuning (a), and detuning (b) of QD-cavity in the case of without pumping (blue) and with pumping (red) process.
	In the non-detuning system, the efficiency is decreased with increasing the temperature (see \fig{fig07}(a)). Increasing the temperature or the pure dephasing leads to decrease in the QD's quality factor, $\mathcal{Q}_{\rm QD} = \omega_{\rm QD}/(\gamma + \gamma^*)$, and the effective quality factor, $\mathcal{Q}_{\rm eff}$, is thus suppressed with increasing $T$, where the $\mathcal{Q}_{\rm eff}$ is defined via

	\begin{equation}
		\frac{1}{\mathcal{Q}_{\rm eff}} = \frac{1}{\mathcal{Q}_{\rm QD}} + \frac{1}{\mathcal{Q}_{\rm c}},
	\end{equation}
	
	with $\mathcal{Q}_{\rm c} = \omega_{\rm c}/\kappa$ the cavity quality factor. The spontaneous emission rate or the effective coupling rate is reduced as it is shown in \fig{fig08} for the non-detuning (a), and detuning (b) of QD-cavity in the case of without pumping (blue) and with pumping (red) process. Consequently, the efficiency is decreased with increasing $T$ in the non-detuning system.
	If the non-dutuning QD-cavity system is pumped, the efficiency is slightly decreased till $T = 100$~K which is caused by decreasing of effective coupling rate shown in \fig{fig08}(a). 
	It should be mentioned that an abrupt drop in the efficiency, and spontaneous emission rate is seen near $T = 100$~K. The abrupt drop is related to the pure dephasing that depends on the interaction of excitons with acoustic and optical phonons of the InGaAs quantum dots.   
	At high temperature above $100$~K, it is natural to interpret the fast dephasing related to 
	the elastic acoustic-phonon scattering. It means that the fast dephasing decays
	from $T > 100$~K to room temperature, $T = 300$~K due to the elastic acoustic-phonon scattering.
	The dephasing decay will strongly influence the efficiency decay shown in \fig{fig07}(a) and (b), 
	and effective transfer rate decay (\fig{fig08}).
	However, the changes of efficiency, and effective transfer rate with temperature compared to the acoustic-phonon band below $100$~K supports the possibility that additional processes, such as interactions with optical phonons, are coming into play.
	So, we have two different regimes of temperature which are low ($T < 100$ K) and high ($T > 100$ K) temperature regimes. These two different regimes gives the abrupt drop in the efficiency, and effective transfer rate vs. temperature curves \cite{PhysRevLett.87.157401, PhysRevLett.85.1516}.

	In the case of detuning QD-cavity system (see \fig{fig07}(b), and \fig{fig08}(b)), the effect of temperature is dramatically different. The decrease in $\mathcal{Q}_{\rm QD}$ and thus $\mathcal{Q}_{\rm eff}$ generated by temperature leads to increase the spontaneous emission rate and also the effective coupling rate as it is clearly seen in \fig{fig08}(b) up to $T = 100$~K. 
	This gives rise an enhancement of efficiency in the same range of temperate. 
	It is very interesting to see that the pumping process will further increase the effective coupling rate and the efficiency, as the temperature is directly proportional to the $R$ for the detuning system.

	\begin{figure}[htb]
		\centering
		\includegraphics[width=0.4\textwidth]{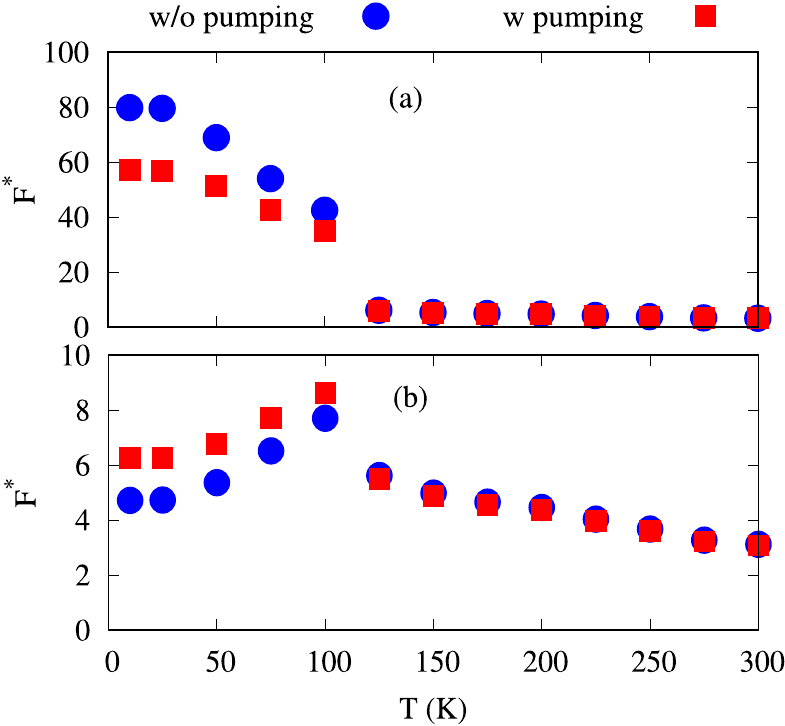}
		\caption{Purcell factor, $F^*$, as a function of temperature for the 
			non-detuning (a) and detuning, ($\delta = 10 \, g$) (b) of QD-cavity system in the case of without pumping, $\mathcal{P}=0.0$ (blue), and with pumping, $\mathcal{P}=2 \, g$ (red), where $\gamma = 0.01g$,
			$g = 50$~$\mu$eV, and $\kappa = 5 \, g$.}
		\label{fig09}
	\end{figure}
	
	We should notice that our model for both non-detuning and detuning QD-cavity system can satisfy the condition of $R \ll k$  in all considered calculations. In this case the system corresponds to the so called Purcell regime. The generalized Purcell factor, $F^*$, can be introduced as $F^* = R/\gamma$ indicating the effective coupling rate is directly proportional to the Purcell factor. 
	The Purcell factor for a detuning system is defined in \eq{Purcell_Factor}, while it is expressed as $F^* = 4 g^2/k \gamma$ for non-detuning system, $\gamma^* = 0.0$ \cite{Gerard_1999, doi:10.1063/1.2964186}. We can thus see that the temperature or pure dephasing has an important role in determining the Purcell factor of a system.

	From experimental point of view for investigating of QD-cavity system, the increasing of the Purcell factor is an influential aim leading to a quest to enhance the $\mathcal{Q}_{\rm c}$, and then efficiency.
	In the case of non-detuning system, $\delta = 0.0$, the Purcell factor is decreased with increasing the temperature or pure dephasing as it is shown in \fig{fig09}(a) for both without pumping (blue) and with pumping (red) of QD-cavity system. This could be expected as the $R$ displayed in \fig{fig08}(a) is also decreased with increasing $T$.  
	In the case of detuning system (see \fig{fig09}(b) ), the Purcell factor follows the same characteristic of the effective coupling strength shown in \fig{fig08}(b). The maximum value of Purcell factor is obtained if $\kappa + \gamma + \gamma^* \approx \delta$ in which the maximum value of the Purcell factor is $F^*_{\rm max} \approx g^2/\gamma \delta$. 
	At $T = 100$~K corresponding to $0.04$ meV of the pure dephasing, $\gamma^*$, we see the maximum value of Purcell factor.
	This is another confirmation of the efficiency enhancement in the detuning system when the temperature is increased.
	
	\begin{figure}[htb]
		\centering
		\includegraphics[width=0.4\textwidth]{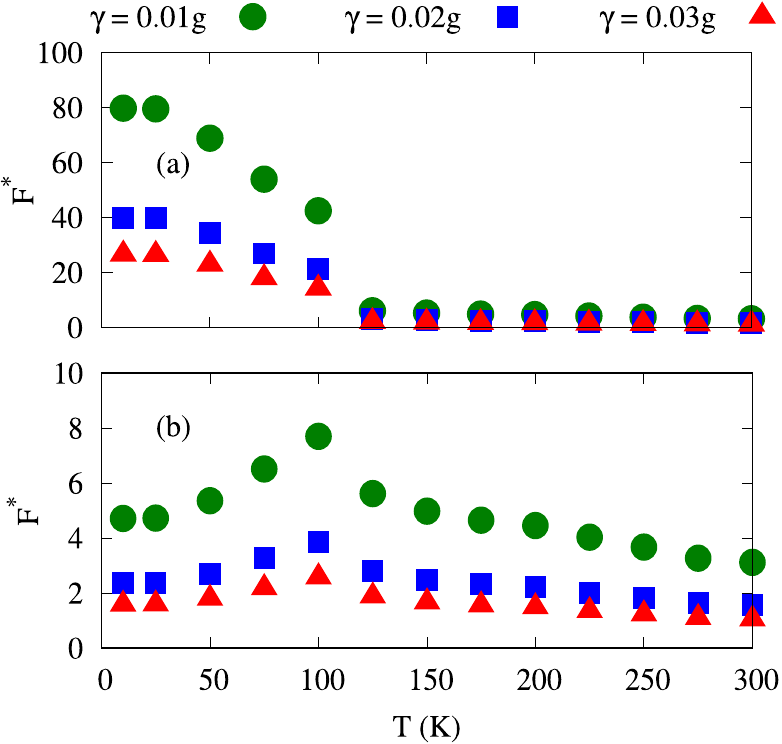}
		\caption{Purcell factor, $F^*$, as a function of temperature for the 
			non-detuning (a) and detuning, ($\delta = 10 \, g$) (b) of QD-cavity system in the case of without pumping, $\mathcal{P}=0.0$ for three different values of $\gamma = 0.01 \, g$ (green), $0.02 \, g$ (blue), and $0.03 \, g$ (red), where $g = 50$~$\mu$eV, and $\kappa = 5 \, g$.}
		\label{fig10}
	\end{figure}
	
	We have mentioned that the Purcell regime considers a very small value of $\gamma$ comparing to other types of loss rate. We now tune the value of $\gamma$ in order to see this effect on Purcell factor.
	In the previous calculation of efficiency, R and Purcell factor, we have considered that $\gamma = 0.02 \, g$, but different values of $\gamma$ are now considered to calculate the Purcell factor. 
	Figure \ref{fig10} displayed the Purcell factor for $\gamma = 0.01 \, g$ (green), $0.02 \, g$ (blue), and $0.03  \, g$ (red) for the non-detuning (a) and detuning (b) system.
	One can clearly see that the Purcell factor is suppressed with increasing $\gamma$. This is the reason for considering a very low value of $\gamma$ of the system. In such low value of $\gamma$, a much higher value of $R$, Purcell factor, and efficiency can be obtained.

	\section{Conclusion}\label{section_conclusion}
	
	In this study, we considered a QD-cavity system as a source of single photon and investigated the physical properties of the single photon such as efficiency and Purcell factor, as we also assumed the cavity to be pumped with a constant strength. We realized that the efficiency of the single photon source is strongly dependent on the temperature, dephasing rate, detuning and pumping conditions. In the non-detuning system, we found that the reduction in the effective coupling strength enhancing the efficiency reduction by raising temperature consequently, the pumping 
	process will further decreases the efficiency and the Purcell factor.
	The study also showed that the pumping mechanism in the detuning system could be used to
	further enhancement of the efficiency. In addition, the temperature $T=100$~K is critical in both systems, however above that temperature the pumping process has no influences any more on the system. 
	
	\section*{Acknowledgment} %
	This work was financially supported by the University of Sulaimani and 
	the Research center of Komar University of Science and Technology. 
	The computations were performed on resources provided by the Division of Computational 
	Nanoscience at the University of Sulaimani.


\begin{thebibliography}{50}
		\expandafter\ifx\csname url\endcsname\relax
		\def\url#1{\texttt{#1}}\fi
		\expandafter\ifx\csname urlprefix\endcsname\relax\def\urlprefix{URL }\fi
		\expandafter\ifx\csname href\endcsname\relax
		\def\href#1#2{#2} \def\path#1{#1}\fi
		
		\bibitem{Reimer2019}
		M.~E. Reimer, C.~Cher, \href{https://doi.org/10.1038/s41566-019-0544-x}{The
			quest for a perfect single-photon source}, Nature Photonics 13~(11) (2019)
		734--736.
		\newblock \href {https://doi.org/10.1038/s41566-019-0544-x}
		{\path{doi:10.1038/s41566-019-0544-x}}.
		\newline\urlprefix\url{https://doi.org/10.1038/s41566-019-0544-x}
		
		\bibitem{10.1117/12.912969}
		S.~Höfling, C.~Schneider, T.~Heindel, M.~Lermer, T.~B. Hoang, J.~Beetz,
		T.~Braun, L.~Balet, N.~Chauvin, L.~Li, S.~Reitzenstein, A.~Fiore, M.~Kamp,
		A.~Forchel, \href{https://doi.org/10.1117/12.912969}{{Single photon sources
				for quantum information applications}}, in: K.~G. Eyink, F.~Szmulowicz, D.~L.
		Huffaker (Eds.), Quantum Dots and Nanostructures: Synthesis,
		Characterization, and Modeling IX, Vol. 8271, International Society for
		Optics and Photonics, SPIE, 2012, pp. 42 -- 49.
		\newblock \href {https://doi.org/10.1117/12.912969}
		{\path{doi:10.1117/12.912969}}.
		\newline\urlprefix\url{https://doi.org/10.1117/12.912969}
		
		\bibitem{Senellart2017}
		P.~Senellart, G.~Solomon, A.~White,
		\href{https://doi.org/10.1038/nnano.2017.218}{High-performance semiconductor
			quantum-dot single-photon sources}, Nature Nanotechnology 12~(11) (2017)
		1026--1039.
		\newblock \href {https://doi.org/10.1038/nnano.2017.218}
		{\path{doi:10.1038/nnano.2017.218}}.
		\newline\urlprefix\url{https://doi.org/10.1038/nnano.2017.218}
		
		\bibitem{walls2007quantum}
		D.~F. Walls, G.~J. Milburn, Quantum optics, Springer Science \& Business Media,
		2007.
		
		\bibitem{Paesani2020}
		S.~Paesani, M.~Borghi, S.~Signorini, A.~Ma{\"i}nos, L.~Pavesi, A.~Laing,
		\href{https://doi.org/10.1038/s41467-020-16187-8}{Near-ideal spontaneous
			photon sources in silicon quantum photonics}, Nature Communications 11~(1)
		(2020) 2505.
		\newblock \href {https://doi.org/10.1038/s41467-020-16187-8}
		{\path{doi:10.1038/s41467-020-16187-8}}.
		\newline\urlprefix\url{https://doi.org/10.1038/s41467-020-16187-8}
		
		\bibitem{Tomm2021}
		N.~Tomm, A.~Javadi, N.~O. Antoniadis, D.~Najer, M.~C. L{\"o}bl, A.~R. Korsch,
		R.~Schott, S.~R. Valentin, A.~D. Wieck, A.~Ludwig, R.~J. Warburton,
		\href{https://doi.org/10.1038/s41565-020-00831-x}{A bright and fast source of
			coherent single photons}, Nature Nanotechnology 16~(4) (2021) 399--403.
		\newblock \href {https://doi.org/10.1038/s41565-020-00831-x}
		{\path{doi:10.1038/s41565-020-00831-x}}.
		\newline\urlprefix\url{https://doi.org/10.1038/s41565-020-00831-x}
		
		\bibitem{PhysRevLett.89.187901}
		A.~Beveratos, R.~Brouri, T.~Gacoin, A.~Villing, J.-P. Poizat, P.~Grangier,
		\href{https://link.aps.org/doi/10.1103/PhysRevLett.89.187901}{Single photon
			quantum cryptography}, Phys. Rev. Lett. 89 (2002) 187901.
		\newblock \href {https://doi.org/10.1103/PhysRevLett.89.187901}
		{\path{doi:10.1103/PhysRevLett.89.187901}}.
		\newline\urlprefix\url{https://link.aps.org/doi/10.1103/PhysRevLett.89.187901}
		
		\bibitem{Heindel_2012}
		T.~Heindel, C.~A. Kessler, M.~Rau, C.~Schneider, M.~Fürst, F.~Hargart, W.-M.
		Schulz, M.~Eichfelder, R.~Ro{\ss}bach, S.~Nauerth, M.~Lermer, H.~Weier,
		M.~Jetter, M.~Kamp, S.~Reitzenstein, S.~Höfling, P.~Michler, H.~Weinfurter,
		A.~Forchel, \href{https://doi.org/10.1088/1367-2630/14/8/083001}{Quantum key
			distribution using quantum dot single-photon emitting diodes in the red and
			near infrared spectral range}, New Journal of Physics 14~(8) (2012) 083001.
		\newblock \href {https://doi.org/10.1088/1367-2630/14/8/083001}
		{\path{doi:10.1088/1367-2630/14/8/083001}}.
		\newline\urlprefix\url{https://doi.org/10.1088/1367-2630/14/8/083001}
		
		\bibitem{michler2000quantum}
		P.~Michler, A.~Kiraz, C.~Becher, W.~Schoenfeld, P.~Petroff, L.~Zhang, E.~Hu,
		A.~Imamoglu, A quantum dot single-photon turnstile device, science 290~(5500)
		(2000) 2282--2285.
		
		\bibitem{doi:10.1063/1.1415346}
		E.~Moreau, I.~Robert, J.~M. Gérard, I.~Abram, L.~Manin, V.~Thierry-Mieg,
		\href{https://doi.org/10.1063/1.1415346}{Single-mode solid-state single
			photon source based on isolated quantum dots in pillar microcavities},
		Applied Physics Letters 79~(18) (2001) 2865--2867.
		\newblock \href {http://arxiv.org/abs/https://doi.org/10.1063/1.1415346}
		{\path{arXiv:https://doi.org/10.1063/1.1415346}}, \href
		{https://doi.org/10.1063/1.1415346} {\path{doi:10.1063/1.1415346}}.
		\newline\urlprefix\url{https://doi.org/10.1063/1.1415346}
		
		\bibitem{PhysRevB.71.241304}
		A.~Kress, F.~Hofbauer, N.~Reinelt, M.~Kaniber, H.~J. Krenner, R.~Meyer,
		G.~B\"ohm, J.~J. Finley,
		\href{https://link.aps.org/doi/10.1103/PhysRevB.71.241304}{Manipulation of
			the spontaneous emission dynamics of quantum dots in two-dimensional photonic
			crystals}, Phys. Rev. B 71 (2005) 241304.
		\newblock \href {https://doi.org/10.1103/PhysRevB.71.241304}
		{\path{doi:10.1103/PhysRevB.71.241304}}.
		\newline\urlprefix\url{https://link.aps.org/doi/10.1103/PhysRevB.71.241304}
		
		\bibitem{Heinze2015}
		D.~Heinze, D.~Breddermann, A.~Zrenner, S.~Schumacher,
		\href{https://doi.org/10.1038/ncomms9473}{A quantum dot single-photon source
			with on-the-fly all-optical polarization control and timed emission}, Nature
		Communications 6~(1) (2015) 8473.
		\newblock \href {https://doi.org/10.1038/ncomms9473}
		{\path{doi:10.1038/ncomms9473}}.
		\newline\urlprefix\url{https://doi.org/10.1038/ncomms9473}
		
		\bibitem{10.1117/12.543168}
		J.-M. Gerard, B.~Gayral, \href{https://doi.org/10.1117/12.543168}{{Toward
				high-efficiency quantum-dot single-photon sources}}, in: D.~L. Huffaker,
		P.~Bhattacharya (Eds.), Quantum Dots, Nanoparticles, and Nanoclusters, Vol.
		5361, International Society for Optics and Photonics, SPIE, 2004, pp. 88 --
		95.
		\newblock \href {https://doi.org/10.1117/12.543168}
		{\path{doi:10.1117/12.543168}}.
		\newline\urlprefix\url{https://doi.org/10.1117/12.543168}
		
		\bibitem{PhysRevB.90.155303}
		K.~H. Madsen, S.~Ates, J.~Liu, A.~Javadi, S.~M. Albrecht, I.~Yeo, S.~Stobbe,
		P.~Lodahl,
		\href{https://link.aps.org/doi/10.1103/PhysRevB.90.155303}{Efficient
			out-coupling of high-purity single photons from a coherent quantum dot in a
			photonic-crystal cavity}, Phys. Rev. B 90 (2014) 155303.
		\newblock \href {https://doi.org/10.1103/PhysRevB.90.155303}
		{\path{doi:10.1103/PhysRevB.90.155303}}.
		\newline\urlprefix\url{https://link.aps.org/doi/10.1103/PhysRevB.90.155303}
		
		\bibitem{Wang2019}
		H.~Wang, Y.-M. He, T.-H. Chung, H.~Hu, Y.~Yu, S.~Chen, X.~Ding, M.-C. Chen,
		J.~Qin, X.~Yang, R.-Z. Liu, Z.-C. Duan, J.-P. Li, S.~Gerhardt, K.~Winkler,
		J.~Jurkat, L.-J. Wang, N.~Gregersen, Y.-H. Huo, Q.~Dai, S.~Yu,
		S.~H{\"o}fling, C.-Y. Lu, J.-W. Pan,
		\href{https://doi.org/10.1038/s41566-019-0494-3}{Towards optimal
			single-photon sources from polarized microcavities}, Nature Photonics 13~(11)
		(2019) 770--775.
		\newblock \href {https://doi.org/10.1038/s41566-019-0494-3}
		{\path{doi:10.1038/s41566-019-0494-3}}.
		\newline\urlprefix\url{https://doi.org/10.1038/s41566-019-0494-3}
		
		\bibitem{Cui_05}
		G.~Cui, M.~G. Raymer,
		\href{http://www.osapublishing.org/oe/abstract.cfm?URI=oe-13-24-9660}{Quantum
			efficiency of single-photon sources in the cavity-QED strong-coupling
			regime}, Opt. Express 13~(24) (2005) 9660--9665.
		\newblock \href {https://doi.org/10.1364/OPEX.13.009660}
		{\path{doi:10.1364/OPEX.13.009660}}.
		\newline\urlprefix\url{http://www.osapublishing.org/oe/abstract.cfm?URI=oe-13-24-9660}
		
		\bibitem{Santori2002}
		C.~Santori, D.~Fattal, J.~Vu{\v{c}}kovi{\'{c}}, G.~S. Solomon, Y.~Yamamoto,
		\href{https://doi.org/10.1038/nature01086}{Indistinguishable photons from a
			single-photon device}, Nature 419~(6907) (2002) 594--597.
		\newblock \href {https://doi.org/10.1038/nature01086}
		{\path{doi:10.1038/nature01086}}.
		\newline\urlprefix\url{https://doi.org/10.1038/nature01086}
		
		\bibitem{PhysRevB.100.155420}
		M.~Reindl, J.~H. Weber, D.~Huber, C.~Schimpf, S.~F. Covre~da Silva, S.~L.
		Portalupi, R.~Trotta, P.~Michler, A.~Rastelli,
		\href{https://link.aps.org/doi/10.1103/PhysRevB.100.155420}{Highly
			indistinguishable single photons from incoherently excited quantum dots},
		Phys. Rev. B 100 (2019) 155420.
		\newblock \href {https://doi.org/10.1103/PhysRevB.100.155420}
		{\path{doi:10.1103/PhysRevB.100.155420}}.
		\newline\urlprefix\url{https://link.aps.org/doi/10.1103/PhysRevB.100.155420}
		
		\bibitem{Dusanowski_2020}
		{\L}.~Dusanowski, D.~K{\"o}ck, E.~Shin, S.-H. Kwon, C.~Schneider,
		S.~H{\"o}fling,
		\href{https://doi.org/10.1021/acs.nanolett.0c01771}{Purcell-enhanced and
			indistinguishable single-photon generation from quantum dots coupled to
			on-chip integrated ring resonators}, Nano Letters 20~(9) (2020) 6357--6363.
		\newblock \href {https://doi.org/10.1021/acs.nanolett.0c01771}
		{\path{doi:10.1021/acs.nanolett.0c01771}}.
		\newline\urlprefix\url{https://doi.org/10.1021/acs.nanolett.0c01771}
		
		\bibitem{Dory2016}
		C.~Dory, K.~A. Fischer, K.~M{\"u}ller, K.~G. Lagoudakis, T.~Sarmiento,
		A.~Rundquist, J.~L. Zhang, Y.~Kelaita, J.~Vu{\v{c}}kovi{\'{c}},
		\href{https://doi.org/10.1038/srep25172}{Complete coherent control of a
			quantum dot strongly coupled to a nanocavity}, Scientific Reports 6~(1)
		(2016) 25172.
		\newblock \href {https://doi.org/10.1038/srep25172}
		{\path{doi:10.1038/srep25172}}.
		\newline\urlprefix\url{https://doi.org/10.1038/srep25172}
		
		\bibitem{Alex_2010}
		A.~Kuhn, D.~Ljunggren†,
		\href{https://doi.org/10.1080/00107511003602990}{Cavity-based single-photon
			sources}, Contemporary Physics 51~(4) (2010) 289--313.
		\newblock \href
		{http://arxiv.org/abs/https://doi.org/10.1080/00107511003602990}
		{\path{arXiv:https://doi.org/10.1080/00107511003602990}}, \href
		{https://doi.org/10.1080/00107511003602990}
		{\path{doi:10.1080/00107511003602990}}.
		\newline\urlprefix\url{https://doi.org/10.1080/00107511003602990}
		
		\bibitem{PhysRevA.79.053838}
		A.~Auff\`eves, J.-M. G\'erard, J.-P. Poizat,
		\href{https://link.aps.org/doi/10.1103/PhysRevA.79.053838}{Pure emitter
			dephasing: A resource for advanced solid-state single-photon sources}, Phys.
		Rev. A 79 (2009) 053838.
		\newblock \href {https://doi.org/10.1103/PhysRevA.79.053838}
		{\path{doi:10.1103/PhysRevA.79.053838}}.
		\newline\urlprefix\url{https://link.aps.org/doi/10.1103/PhysRevA.79.053838}
		
		\bibitem{PhysRevB.75.073308}
		I.~Favero, A.~Berthelot, G.~Cassabois, C.~Voisin, C.~Delalande, P.~Roussignol,
		R.~Ferreira, J.~M. G\'erard,
		\href{https://link.aps.org/doi/10.1103/PhysRevB.75.073308}{Temperature
			dependence of the zero-phonon linewidth in quantum dots: An effect of the
			fluctuating environment}, Phys. Rev. B 75 (2007) 073308.
		\newblock \href {https://doi.org/10.1103/PhysRevB.75.073308}
		{\path{doi:10.1103/PhysRevB.75.073308}}.
		\newline\urlprefix\url{https://link.aps.org/doi/10.1103/PhysRevB.75.073308}
		
		\bibitem{RevModPhys.87.1379}
		A.~Reiserer, G.~Rempe,
		\href{https://link.aps.org/doi/10.1103/RevModPhys.87.1379}{Cavity-based
			quantum networks with single atoms and optical photons}, Rev. Mod. Phys. 87
		(2015) 1379--1418.
		\newblock \href {https://doi.org/10.1103/RevModPhys.87.1379}
		{\path{doi:10.1103/RevModPhys.87.1379}}.
		\newline\urlprefix\url{https://link.aps.org/doi/10.1103/RevModPhys.87.1379}
		
		\bibitem{PhysRevB.81.245419}
		A.~Auff\`eves, D.~Gerace, J.-M. G\'erard, M.~F. m.~c. Santos, L.~C. Andreani,
		J.-P. Poizat,
		\href{https://link.aps.org/doi/10.1103/PhysRevB.81.245419}{Controlling the
			dynamics of a coupled atom-cavity system by pure dephasing}, Phys. Rev. B 81
		(2010) 245419.
		\newblock \href {https://doi.org/10.1103/PhysRevB.81.245419}
		{\path{doi:10.1103/PhysRevB.81.245419}}.
		\newline\urlprefix\url{https://link.aps.org/doi/10.1103/PhysRevB.81.245419}
		
		\bibitem{ABDULLAH2020114221}
		N.~R. Abdullah, C.-S. Tang, A.~Manolescu, V.~Gudmundsson,
		\href{https://www.sciencedirect.com/science/article/pii/S1386947719311750}{Oscillations
			in electron transport caused by multiple resonances in a quantum dot-qed
			system in the steady-state regime}, Physica E: Low-dimensional Systems and
		Nanostructures 123 (2020) 114221.
		\newblock \href {https://doi.org/https://doi.org/10.1016/j.physe.2020.114221}
		{\path{doi:https://doi.org/10.1016/j.physe.2020.114221}}.
		\newline\urlprefix\url{https://www.sciencedirect.com/science/article/pii/S1386947719311750}
		
		\bibitem{JOHANSSON20121760}
		J.~Johansson, P.~Nation, F.~Nori,
		\href{https://www.sciencedirect.com/science/article/pii/S0010465512000835}{Qutip:
			An open-source python framework for the dynamics of open quantum systems},
		Computer Physics Communications 183~(8) (2012) 1760--1772.
		\newblock \href {https://doi.org/https://doi.org/10.1016/j.cpc.2012.02.021}
		{\path{doi:https://doi.org/10.1016/j.cpc.2012.02.021}}.
		\newline\urlprefix\url{https://www.sciencedirect.com/science/article/pii/S0010465512000835}
		
		\bibitem{JOHANSSON20131234}
		J.~Johansson, P.~Nation, F.~Nori,
		\href{https://www.sciencedirect.com/science/article/pii/S0010465512003955}{Qutip
			2: A python framework for the dynamics of open quantum systems}, Computer
		Physics Communications 184~(4) (2013) 1234--1240.
		\newblock \href {https://doi.org/https://doi.org/10.1016/j.cpc.2012.11.019}
		{\path{doi:https://doi.org/10.1016/j.cpc.2012.11.019}}.
		\newline\urlprefix\url{https://www.sciencedirect.com/science/article/pii/S0010465512003955}
		
		\bibitem{PhysRevB.82.195325}
		N.~R. Abdullah, C.-S. Tang, V.~Gudmundsson,
		\href{http://link.aps.org/doi/10.1103/PhysRevB.82.195325}{{Time-dependent
				magnetotransport in an interacting double quantum wire with window
				coupling}}, Phys. Rev. B 82 (2010) 195325.
		\newblock \href {https://doi.org/10.1103/PhysRevB.82.195325}
		{\path{doi:10.1103/PhysRevB.82.195325}}.
		\newline\urlprefix\url{http://link.aps.org/doi/10.1103/PhysRevB.82.195325}
		
		\bibitem{doi:10.1021/acsphotonics.5b00115}
		V.~Gudmundsson, A.~Sitek, P.-y. Lin, N.~R. Abdullah, C.-S. Tang, A.~Manolescu,
		\href{https://doi.org/10.1021/acsphotonics.5b00115}{Coupled collective and
			rabi oscillations triggered by electron transport through a photon cavity},
		ACS Photonics 2~(7) (2015) 930--934.
		\newblock \href
		{http://arxiv.org/abs/https://doi.org/10.1021/acsphotonics.5b00115}
		{\path{arXiv:https://doi.org/10.1021/acsphotonics.5b00115}}, \href
		{https://doi.org/10.1021/acsphotonics.5b00115}
		{\path{doi:10.1021/acsphotonics.5b00115}}.
		\newline\urlprefix\url{https://doi.org/10.1021/acsphotonics.5b00115}
		
		\bibitem{Vidar:ANDP201500298}
		V.~Gudmundsson, A.~Sitek, N.~R. Abdullah, C.-S. Tang, A.~Manolescu,
		\href{http://dx.doi.org/10.1002/andp.201500298}{{Cavity-photon contribution
				to the effective interaction of electrons in parallel quantum dots}}, Annalen
		der Physik 528~(5) (2016) 394--403.
		\newblock \href {https://doi.org/10.1002/andp.201500298}
		{\path{doi:10.1002/andp.201500298}}.
		\newline\urlprefix\url{http://dx.doi.org/10.1002/andp.201500298}
		
		\bibitem{doi:10.1063/1.3527930}
		V.~Loo, L.~Lanco, A.~Lemaître, I.~Sagnes, O.~Krebs, P.~Voisin, P.~Senellart,
		\href{https://doi.org/10.1063/1.3527930}{Quantum dot-cavity strong-coupling
			regime measured through coherent reflection spectroscopy in a very high-Q
			micropillar}, Applied Physics Letters 97~(24) (2010) 241110.
		\newblock \href {http://arxiv.org/abs/https://doi.org/10.1063/1.3527930}
		{\path{arXiv:https://doi.org/10.1063/1.3527930}}, \href
		{https://doi.org/10.1063/1.3527930} {\path{doi:10.1063/1.3527930}}.
		\newline\urlprefix\url{https://doi.org/10.1063/1.3527930}
		
		\bibitem{englund2009coherent}
		D.~Englund, A.~Majumdar, A.~Faraon, M.~Toishi, N.~Stoltz, P.~Petroff,
		J.~Vuckovic, Coherent excitation of a strongly coupled quantum dot-cavity
		system, arXiv preprint arXiv:0902.2428 (2009).
		
		\bibitem{https://doi.org/10.1002/lpor.200810081}
		P.~Yao, V.~Manga~Rao, S.~Hughes,
		\href{https://onlinelibrary.wiley.com/doi/abs/10.1002/lpor.200810081}{On-chip
			single photon sources using planar photonic crystals and single quantum
			dots}, Laser \& Photonics Reviews 4~(4) (2010) 499--516.
		\newblock \href
		{http://arxiv.org/abs/https://onlinelibrary.wiley.com/doi/pdf/10.1002/lpor.200810081}
		{\path{arXiv:https://onlinelibrary.wiley.com/doi/pdf/10.1002/lpor.200810081}},
		\href {https://doi.org/https://doi.org/10.1002/lpor.200810081}
		{\path{doi:https://doi.org/10.1002/lpor.200810081}}.
		\newline\urlprefix\url{https://onlinelibrary.wiley.com/doi/abs/10.1002/lpor.200810081}
		
		\bibitem{JONSSON201781}
		T.~H. Jonsson, A.~Manolescu, H.-S. Goan, N.~R. Abdullah, A.~Sitek, C.-S. Tang,
		V.~Gudmundsson,
		\href{http://www.sciencedirect.com/science/article/pii/S001046551730200X}{{Efficient
				determination of the Markovian time-evolution towards a steady-state of a
				complex open quantum system}}, Computer Physics Communications 220 (2017)
		81--90.
		\newblock \href {https://doi.org/10.1016/j.cpc.2017.06.018}
		{\path{doi:10.1016/j.cpc.2017.06.018}}.
		\newline\urlprefix\url{http://www.sciencedirect.com/science/article/pii/S001046551730200X}
		
		\bibitem{7798963}
		R.~Azouit, A.~Sarlette, P.~Rouchon, Adiabatic elimination for open quantum
		systems with effective lindblad master equations, in: 2016 IEEE 55th
		Conference on Decision and Control (CDC), 2016, pp. 4559--4565.
		\newblock \href {https://doi.org/10.1109/CDC.2016.7798963}
		{\path{doi:10.1109/CDC.2016.7798963}}.
		
		\bibitem{Azouit_2017}
		R.~Azouit, F.~Chittaro, A.~Sarlette, P.~Rouchon,
		\href{https://doi.org/10.1088/2058-9565/aa7f3f}{Towards generic adiabatic
			elimination for bipartite open quantum systems}, Quantum Science and
		Technology 2~(4) (2017) 044011.
		\newblock \href {https://doi.org/10.1088/2058-9565/aa7f3f}
		{\path{doi:10.1088/2058-9565/aa7f3f}}.
		\newline\urlprefix\url{https://doi.org/10.1088/2058-9565/aa7f3f}
		
		\bibitem{GUDMUNDSSON20181672}
		V.~Gudmundsson, N.~R. Abdullah, A.~Sitek, H.-S. Goan, C.-S. Tang, A.~Manolescu,
		\href{http://www.sciencedirect.com/science/article/pii/S0375960118303748}{{Current
				correlations for the transport of interacting electrons through parallel
				quantum dots in a photon cavity}}, Physics Letters A 382~(25) (2018)
		1672--1678.
		\newblock \href {https://doi.org/10.1016/j.physleta.2018.04.017}
		{\path{doi:10.1016/j.physleta.2018.04.017}}.
		\newline\urlprefix\url{http://www.sciencedirect.com/science/article/pii/S0375960118303748}
		
		\bibitem{PhysRevLett.87.157401}
		P.~Borri, W.~Langbein, S.~Schneider, U.~Woggon, R.~L. Sellin, D.~Ouyang,
		D.~Bimberg,
		\href{https://link.aps.org/doi/10.1103/PhysRevLett.87.157401}{Ultralong
			dephasing time in InGaAs quantum dots}, Phys. Rev. Lett. 87 (2001) 157401.
		\newblock \href {https://doi.org/10.1103/PhysRevLett.87.157401}
		{\path{doi:10.1103/PhysRevLett.87.157401}}.
		\newline\urlprefix\url{https://link.aps.org/doi/10.1103/PhysRevLett.87.157401}
		
		\bibitem{PhysRevLett.103.087405}
		A.~Laucht, N.~Hauke, J.~M. Villas-B\^oas, F.~Hofbauer, G.~B\"ohm, M.~Kaniber,
		J.~J. Finley,
		\href{https://link.aps.org/doi/10.1103/PhysRevLett.103.087405}{Dephasing of
			exciton polaritons in photoexcited InGaAs quantum dots in GaAs nanocavities},
		Phys. Rev. Lett. 103 (2009) 087405.
		\newblock \href {https://doi.org/10.1103/PhysRevLett.103.087405}
		{\path{doi:10.1103/PhysRevLett.103.087405}}.
		\newline\urlprefix\url{https://link.aps.org/doi/10.1103/PhysRevLett.103.087405}
		
		\bibitem{doi:10.1021/acs.nanolett.5b03724}
		T.~B. Hoang, G.~M. Akselrod, M.~H. Mikkelsen,
		\href{https://doi.org/10.1021/acs.nanolett.5b03724}{Ultrafast
			room-temperature single photon emission from quantum dots coupled to
			plasmonic nanocavities}, Nano Letters 16~(1) (2016) 270--275, pMID: 26606001.
		\newblock \href
		{http://arxiv.org/abs/https://doi.org/10.1021/acs.nanolett.5b03724}
		{\path{arXiv:https://doi.org/10.1021/acs.nanolett.5b03724}}, \href
		{https://doi.org/10.1021/acs.nanolett.5b03724}
		{\path{doi:10.1021/acs.nanolett.5b03724}}.
		\newline\urlprefix\url{https://doi.org/10.1021/acs.nanolett.5b03724}
		
		\bibitem{Ishida2013}
		N.~Ishida, T.~Byrnes, F.~Nori, Y.~Yamamoto,
		\href{https://doi.org/10.1038/srep01180}{Photoluminescence of a microcavity
			quantum dot system in the quantum strong-coupling regime}, Scientific Reports
		3~(1) (2013) 1180.
		\newblock \href {https://doi.org/10.1038/srep01180}
		{\path{doi:10.1038/srep01180}}.
		\newline\urlprefix\url{https://doi.org/10.1038/srep01180}
		
		\bibitem{PhysRevLett.85.1516}
		A.~V. Uskov, A.-P. Jauho, B.~Tromborg, J.~M\o{}rk, R.~Lang,
		\href{https://link.aps.org/doi/10.1103/PhysRevLett.85.1516}{Dephasing times
			in quantum dots due to elastic lo phonon-carrier collisions}, Phys. Rev.
		Lett. 85 (2000) 1516--1519.
		\newblock \href {https://doi.org/10.1103/PhysRevLett.85.1516}
		{\path{doi:10.1103/PhysRevLett.85.1516}}.
		\newline\urlprefix\url{https://link.aps.org/doi/10.1103/PhysRevLett.85.1516}
		
		\bibitem{Gerard_1999}
		J.-M. Gerard, B.~Gayral, Strong purcell effect for InAs quantum boxes in
		three-dimensional solid-state microcavities, Journal of Lightwave Technology
		17~(11) (1999) 2089--2095.
		\newblock \href {https://doi.org/10.1109/50.802999}
		{\path{doi:10.1109/50.802999}}.
		
		\bibitem{doi:10.1063/1.2964186}
		M.~Francardi, L.~Balet, A.~Gerardino, N.~Chauvin, D.~Bitauld, L.~H. Li,
		B.~Alloing, A.~Fiore, \href{https://doi.org/10.1063/1.2964186}{Enhanced
			spontaneous emission in a photonic-crystal light-emitting diode}, Applied
		Physics Letters 93~(14) (2008) 143102.
		\newblock \href {http://arxiv.org/abs/https://doi.org/10.1063/1.2964186}
		{\path{arXiv:https://doi.org/10.1063/1.2964186}}, \href
		{https://doi.org/10.1063/1.2964186} {\path{doi:10.1063/1.2964186}}.
		\newline\urlprefix\url{https://doi.org/10.1063/1.2964186}
		
	\end{thebibliography}

\end{document}